\documentclass[12pt]{iopart}
\usepackage{graphicx}
\begin{document}

\title[Channeling of Protons]{Channeling of Protons Through Carbon Nanotubes Embedded in Dielectric Media}

\author{D. Borka$^{1,2}$, D. J. Mowbray$^2$, Z. L. Mi\v{s}kovi\'{c}$^2$, S. Petrovi\'{c}$^1$ and N. Ne\v{s}kovi\'{c}$^1$}

\address{$^1$Laboratory of Physics (010), Vin\v{c}a Institute of Nuclear Sciences, P.O. Box 522\\
11001 Belgrade, Serbia}
\address{$^2$Department of Applied Mathematics, University of Waterloo\\
Waterloo, Ontario, Canada N2L3G1}
\ead{dusborka@vin.bg.ac.yu}

\begin{abstract}
We investigate how the dynamic polarization of the carbon atoms
valence electrons affects the spatial   distributions of protons
channeled in the (11, 9) single-wall carbon nanotubes placed in
vacuum and embedded in various dielectric media. The initial proton
speed is varied between 3 and 8 a.u., corresponding to the energies
between 0.223 and 1.59 MeV, respectively, while the nanotube length
is varied between 0.1 and 0.8 $\mu$m. The spatial distributions of
channeled protons are generated using a computer simulation method,
which includes the numerical solving of the proton equations of
motion in the transverse plane. We show that the dynamic
polarization effect can strongly affect the rainbow maxima in the
spatial distributions, so as to increase the proton flux at the
distances from the nanotube wall of the order of a few tenths of a
nanometer at the expense of the flux at the nanotube center. While
our findings are connected to the possible applications of nanosized
ion beams created with the nanotubes embedded in various dielectric
media for biomedical research and in materials modification, they
also open the prospects of applying ion channeling for detecting and
locating the atoms and molecules intercalated inside the nanotubes.
\end{abstract}

\vspace{2pc}


\maketitle

\section{Introduction}

Theoretical modeling of ion channeling through carbon nanotubes has
reached a mature level \cite{mour07mour05,artr05,misk07} whereas
experimental realization of this process is still at the preliminary
stage \cite{zhu05,chai07}. Nevertheless, theoreticians continue to
explore possible applications of ion channeling through the
nanotubes, beginning with those based on the \emph{angular}
distributions of channeled ions, e.g., for the purpose of deflecting
ion beams in accelerators \cite{biry05a,biry05b} or determining some
structural details of the short nanotubes using the rainbow effect
\cite{petr05a,petr05b,borka05,nesk05}. Besides, theoretical studies
of the \emph{spatial} distributions of channeled ions have
demonstrated the possibility of creating nanosized ion beams, which
could find interesting applications in biomedical research and for
materials modification \cite{biry05a,biry05c,bell03,bell05,zhou05}.
However, an additional application of ion channeling through the
nanotubes based on the \emph{spatial} distributions remains to be
investigated in the context of extending the classical use of ion
channeling for materials analysis \cite{feld82} to the nanotube
based materials.

Namely, one of the most powerful applications of ion channeling
through single crystals was developed to detect and locate their
defects. It is based on measuring ion dechanneling on such defects
in combination with their Rutherford backscattering \cite{feld82}.
It was crucial for this application of ion channeling to develop the
adequate theoretical models for calculating the spatial
distributions of channeled ions, and thus enable the precise control
and interpretation of the measurements, providing the information on
such defects \cite{feld82}. On the other hand, the nanotubes can
exhibit numerous defects, appearing during their synthesis or due to
external agents such as particle irradiation \cite{kota07}.
Moreover, one of the main directions within the nanotube research is
related to the phenomena of intercalation of the atomic and
molecular species in them \cite{ducl02}. Therefore, it is reasonable
to explore whether ion channeling can be also used to analyze
defects in the nanotubes, in particular, to detect and locate the
atoms and molecules intercalated in them. In a first attempt to
address this problem, we present here a detailed study of the
spatial distributions of protons channeled through the nanotubes,
with special emphasis on the rainbow effect in the transverse plane
with its possible applications.

Having in mind that the ion channeling regime is most suitable for
probing the atoms intercalated in the nanotubes involves light ions
in the MeV energy range \cite{feld82}, one must emphasize that, for
such projectiles, the dynamic polarization of the nanotube atoms
valence electrons can give rise to the strong image force, which can
pull the channeled ions towards the nanotube wall
\cite{mowb04mowb04}. Of course, one expects that the intercalated
atoms would be placed right there, close to the nanotube wall, in
some state of adsorption on the nanotube atoms surface. While the
image force has not been found to play any significant role in
channels of single crystals \cite{feld82}, its role has been
identified clearly in ion-surface scattering \cite{wint02}, and in
ion transmission through capillaries in solids
\cite{aris01aris01,toke05}. In addition, although the image force
plays the minor roles in ion channeling through the nanotubes in the
GeV and keV energy ranges
\cite{mour07mour05,artr05,biry05a,biry05b,kras05}, it has been shown
to influence strongly ion trajectories in the MeV energy range. For
example, the image force gives rise to the rainbow effect in angular
distributions of protons channeled through the short chiral
single-wall and double wall nanotubes \cite{borka06,borka07}.

It has been established that, when the nanotubes are grown in a
dielectric medium, one can achieve a very high degree of their
ordering and straightening. This makes such composite structures
very suitable candidates to be used in experiments of ion channeling
through the nanotubes. Therefore, it is important to study the
effects of the surrounding dielectric medium on ion channeling
through the nanotubes in the MeV energy range. The materials of
interest are Al$_2$O$_3$ \cite{zhu05}, SiO$_2$ \cite{berd,tset06} as
well as Ni \cite{guer94} and Pt \cite{chai07}. In this context, it
has been shown recently that the image force and, consequently, the
rainbow effect in the angular distributions of protons channeled
through the nanotubes can be strongly modified by the polarization
of the surrounding cylindrical dielectric boundary
\cite{mowb06mowb07,borka08}. The separation between the nanotube and
the SiO$_2$ and Al$_2$O$_3$ boundaries is approximated by the
nanotube atom van der Waals radius \cite{hulm04,hulm03}. The
separation between the nanotube and the Ni boundary is approximated
by the separation between the a graphene and a (111) Ni surface,
which was calculated using a minimization structural calculation
based on the density functional theory \cite{sole02}. In both cases
the obtained separation was 3.21 a.u.

We shall investigate here how the spatial distributions of protons
channeled in the (11, 9) single-wall nanotubes placed in vacuum or
embedded in various dielectric media are affected by the dynamic
polarization effect. While such a study complements our previous
studies of the influence of the image force on the rainbow effect in
the angular distributions of protons channeled in the nanotubes
\cite{borka06,borka07,borka08}, we expect that the analogous
influence on the spatial distribution of channeled protons may lead
to a significant change of the proton flux within the nanotube,
enabling one to probe the atoms and molecules adsorbed on the
nanotube wall. The image force acting on the proton moving through
the nanotube embedded in a dielectric medium will be calculated
using a two-dimensional hydrodynamic model of the nanotube atoms
valence electrons while the surrounding dielectric medium will be
described by a suitable dielectric function \cite{mowb06mowb07}.

After outlining the basic theory used in modeling the interaction
potentials between the proton and the nanotube and dielectric
medium, we shall discuss the results of the proton trajectory
simulations and give the concluding remarks.

The atomic units will be used throughout the paper unless explicitly
stated otherwise.

\section{Theory}

The system under investigation is a proton moving through an (11, 9)
single-wall carbon nanotube embedded in a dielectric medium. The $z$
axis coincides with the nanotube axis and the origin lies in its
entrance transverse plane. The initial proton speed vector is taken
to be parallel to the $z$ axis. We assume that the nanotube is
sufficiently short for the proton energy loss to be neglected. The
initial proton speed, $v$, is varied between 3 and 8 a.u.,
corresponding to the energies between 0.223 and 1.59 MeV,
respectively. The nanotube length, $L$, is varied between 0.1 and
0.8 $\mu$m.

We assume that the (repulsive) interaction between the proton and
nanotube atoms can be treated classically using the Doyle-Turner
expression for the proton-nanotube atom interaction potential
averaged axially and azimuthally \cite{lind65,zhev98zhev00,doyl68}.
The resulting interaction potential reads

\begin{equation}
U_{rep}(r) = \frac{16\pi Z_1 Z_2 d}{3\sqrt 3 l^2 }\sum\limits_{j =
1}^4 {a_j b_j^2 I_0 (b_j^2 rd)\exp \{- b_j^2 } [r^2 + (d/2)^2 ]\},
\label{equ01}
\end{equation}

\noindent where $Z_{1}$ = 1 and $Z_{2}$ = 6 are the atomic numbers
of the proton and nanotube atom, respectively, $d$ is the nanotube
diameter, $l$ is the nanotube atoms bond length, $r$ is the distance
between the proton and nanotube axis, $I_0$ designate the modified
Bessel function of the first kind and the $0^{th}$ order, and $a_j =
(0.115, 0.188, 0.072, 0.020)$ and $b_j = (0.547, 0.989, 1.982,
5.656)$ are the fitting parameters \cite{doyl68}.

The dynamic polarization of the nanotube and dielectric medium by
the proton is treated using a two-dimensional hydrodynamic model of
the nanotube atoms valence electrons, based on a jellium-like
description of the nanotube ion cores, extended to include the
contribution of the dielectric boundary
\cite{doer04,mowb04mowb04,mowb06mowb07}. This model includes the
procedures of axial and azimuthal averaging, as the repulsive
interaction model. It gives the (attractive) interaction potential
between the proton and its image. Let us designate the proton
position at time $t$ by $\vec r = \vec r(t)$ and the electric
potential at point $\vec R$ originating from the (screened) proton,
perturbing the nanotube and dielectric boundary, by $\Phi_{ext}(\vec
R, \vec r, t)$. This potential is called the external potential. The
image interaction potential at the proton position reads

\begin{equation}
U_{atr} = - \frac{Z_1}{2}\Phi_{ind} ( \vec r,t ),
\label{equ02}
\end{equation}

\noindent where $\Phi_{ind} (\vec r,t)$ is the electric potential at
the proton position originating from the polarization charges
induced on the nanotube and dielectric boundary by the (screened)
proton. This potential is called the induced potential.

Since the repulsive and attractive interaction potentials are
axially symmetric, the initial proton speed vector is parallel to
the $z$ axis and the proton is channeled, the cylindrical
coordinates of the proton position are $r = r(t)$, $\varphi =
\varphi_0$ and $z = vt$, where $\varphi_0$ is the azimuthal
coordinate of the initial proton position. Let us designate the
nanotube and dielectric boundary radii by $a = d/2$ and $b$,
respectively. Then, the Fourier transform of the external potential
at the nanotube or dielectric boundary, i.e., for $R = a$ or $R =
b$, is

\begin{equation}
\tilde \Phi_{ext}(R,r) = \frac{2\pi
Z_1}{\varepsilon_{nt}}g(R,r,m,k)\delta(\omega -kv)\exp(-im\varphi_0)
\label{equ03}
\end{equation}

\noindent where $\varepsilon_{nt} = 1$ is the dielectric constant of
the nanotube background, $m$, $k$ and $\omega$ are the angular
oscillation mode, longitudinal wave number and angular frequency of
an elementary excitation of the nanotube atoms valence electrons
treated as an electron gas, and $g(R,r,m,k)$ is the radial Green's
function.

The Fourier transform of the induced potential at the proton
position is

\begin{eqnarray}
\tilde \Phi_{ind}(r) = \frac{-a\tilde
n_a}{\varepsilon_{nt}}[g(a,r,m,k) & + & b g(b,r,m,k)\Re g'(b,a,m,k)] \nonumber\\
& + & b g(b,r,m,k)\Re {\tilde \Phi_{ext}'(b,r)},
\label{equ04}
\end{eqnarray}

\noindent where

\begin{equation}
\tilde n_a = \frac{{\tilde \Phi_{ext}(a,r) + bg(b,a,m,k)\Re {\tilde
\Phi_{ext}'(b,r)}}}{{\chi^{-1} + (a/ \varepsilon_{nt}) [g(a,a,m,k) +
b g(b,a,m,k)\Re g'(b,a,m,k)]}},
\label{equ05}
\end{equation}

\begin{equation}
\chi = \frac{{n_0 (k^2 + {{m^2 } \mathord{\left/
 {\vphantom {{m^2 } {a^2 )}}} \right.
 \kern-\nulldelimiterspace} {a^2 )}}}}{{\alpha (k^2 + {m^2 \mathord{\left/
 {\vphantom {{m^2 } {a^2 )}}} \right.
 \kern-\nulldelimiterspace} {a^2 )}} + \beta (k^2 + {m^2 \mathord{\left/
 {\vphantom {{m^2 } {a^2 )}}} \right.
 \kern-\nulldelimiterspace} {a^2 )}}^2 - \omega (\omega + i\gamma )}},
\label{equ06}
\end{equation}

\noindent and

\begin{equation}
\Re = \frac{\varepsilon_{\omega} - \varepsilon_{nt}}{4\pi
[\varepsilon_{nt} + (\varepsilon_{nt} - \varepsilon_{\omega})kbI_m
(kb)K'_m (kb)]}; \label{equ07}
\end{equation}

\noindent $\tilde n_a$ is the Fourier transform of the surface
polarization charge density induced at the nanotube, $\chi$ is the
response function of the polarization charge induced at the
nanotube, $n_0$ = 0.428 a.u. is the equilibrium density of the
electron gas, $\alpha = \pi n_0$, $\beta = 1/4$, and $\gamma \to
0^+$, $\Re$ is the response function of the polarization charge
induced at the dielectric boundary, $\varepsilon_{\omega}$ is the
dielectric function of the dielectric medium, and $I_m$ and $K_m$
designate the modified Bessel functions of the first and second
kinds and the $m^{th}$ order, respectively; the derivatives of
functions $g(R,r,m,k)$ and $\tilde \Phi_{ext} (R,r)$ are taken with
respect to their first arguments.

The induced potential at the proton position is

\begin{equation}
\Phi_{ind} (\vec r,t) = \sum\limits_{m = - \infty }^{m = \infty}
{\int\limits_{-\infty}^\infty {\frac{dk}{(2 \pi)^2}} \int\limits_{-
\infty}^\infty {\frac{d\omega}{2 \pi}\exp \left\{ {i\left[
{m\varphi_0 - (\omega  - kv)t} \right]} \right\}\tilde \Phi_{ind}(r)
} }. \label{equ08}
\end{equation}

The total interaction potential between the proton and the nanotube
and dielectric medium reads

\begin{equation}
U(\vec r,t) = U_{rep} (r) - \frac{Z_1}{2}\Phi_{ind}  (\vec r,t).
\label{equ09}
\end{equation}

The spatial distributions of channeled protons in the exit
transverse plane are generated using a computer simulation method,
which includes the numerical solving of the proton equations of
motion in the transverse plane. The rectangular coordinates of the
initial proton position, $x_0$ and $y_0$, are chosen randomly from a
two-dimensional uniform distribution with condition $r_{\mathrm{0}}
= ({x_0^2 + y_0^2})^{\frac {1}{2}} < a - a_{sc}$, where $a_{sc} =
\left[ {9\pi^2 /(128 \ Z_{2} )} \right]^{\frac{1}{3}}a_0$ is the
nanotube atom screening radius and $a_0$ the Bohr radius. We take
for the nanotube atoms bond length $l$ = 0.144 nm [38] and obtain
for the nanotube radius $a$ = 0.689 nm. The initial number of
protons is 3~141~929.

It has been demonstrated that proton channeling in nanotubes can be
analyzed successfully via the corresponding mapping of the impact
parameter plane to the scattering angle plane
\cite{petr05a,petr05b,borka05,nesk05,borka06,borka07}. Analogously,
we analyze here the mapping of the entrance transverse plane to the
exit transverse plane. However, since in the case we investigate the
total interaction potential is axially symmetric, the analysis of
this mapping can be reduced to the analysis of the mapping of the
proton radial axis in the entrance transverse plane to the proton
radial axis in the exit transverse plane. Further, we can take that
$y_0$ = 0 and analyze only the mapping of the $x_0$ axis in the
entrance transverse plane to the $x_0$ axis in the exit transverse
plane. The extrema in this mapping are the rainbow maxima or minima,
and the corresponding singularities in the spatial distribution of
channeled protons are the rainbow singularities.

\section{Results and discussion}

We shall analyze first the spatial distributions of protons
channeled in the (11, 9) carbon nanotubes placed in vacuum. The
proton speed will be $v$ = 3 a.u. and the nanotube lengths $L$ =
0.1, 0.2 and 0.3 $\mu$m. For this proton speed the spatial
distributions of channeled protons for the nanotubes embedded in a
dielectric medium do not differ from the corresponding spatial
distributions for the nanotubes placed in vacuum. This part of the
study represents a continuation of our previous study of the
influence of the dynamic polarization effect on the angular
distributions of protons channeled in the (11, 9) nanotubes placed
in vacuum \cite{borka06}.

After that, we shall explore the spatial distributions of protons
channeled in the (11, 9) nanotubes embedded in SiO$_2$. The proton
speed will be $v$ = 5 a.u. and the nanotube lengths $L$ = 0.3 and
0.5 $\mu$m. For this proton speed the spatial distributions of
channeled protons for the nanotubes embedded in a dielectric medium
differ from the corresponding spatial distributions for the
nanotubes placed in vacuum. This part of the study represents a
continuation of our previous study of the influence of the image
interaction potential on the angular distributions of protons
channeled in the (11, 9) nanotubes placed in SiO$_2$ \cite{borka08}.
Then, we shall consider the spatial distributions of protons
channeled in the (11, 9) nanotubes embedded in SiO$_2$, Al$_2$O$_3$
and Ni. The proton speed will be $v$ = 8 a.u. and the nanotube
length $L$ = 0.8 a.u.

The nanotube radius is $a$ = 13.01 a.u. The separation between the
nanotube and the dielectric boundaries is 3.21 a.u. Thus, the
dielectric boundary radius is $b = a$ + 3.21 a.u. We describe the
surrounding SiO$_2$ by the dielectric constant of 3.9 \cite{swar00}.
The dielectric responses of the surrounding Al$_2$O$_3$ and Ni are
modeled using the methods described by Arista and Fuentes
\cite{aris01aris01} and Kwei et al. \cite{kwei93}, respectively. We
have found that in the cases of SiO$_2$ and Al$_2$O$_3$ a proton
moving at a speed below about 3 a.u. does not polarize the
dielectric media. This means that the dielectric medium is
completely screened by the nanotube and does not contribute to the
image force. For a proton speed above about 3 a.u. the polarization
of the dielectric media occurs. Consequently, the screening of the
dielectric medium by the nanotube is incomplete and it influences
the image force.

Figure 1(a) shows two spatial distributions of channeled protons
along the $x$ axis in the exit transverse plane for the proton speed
$v$ = 3 a.u. and the nanotube length $L$ = 0.1 $\mu$m. The nanotube
is placed in vacuum. The spatial distributions correspond to the
cases without the dynamic polarization effect and with it. In each
case the spatial distribution contains a strong central maximum and
a pair of prominent peripheral maxima, designated by 1 in the former
case and by 1$_i$ in the latter case. In the case with the dynamic
polarization effect the central maximum is narrower and weaker than
in the case without it. In the former case the peripheral maxima are
located at $x = \pm$8.4 a.u., and in the latter case at $x =
\pm$10.2 a.u.

\begin{figure}
\centering
\includegraphics[width=0.49\textwidth]{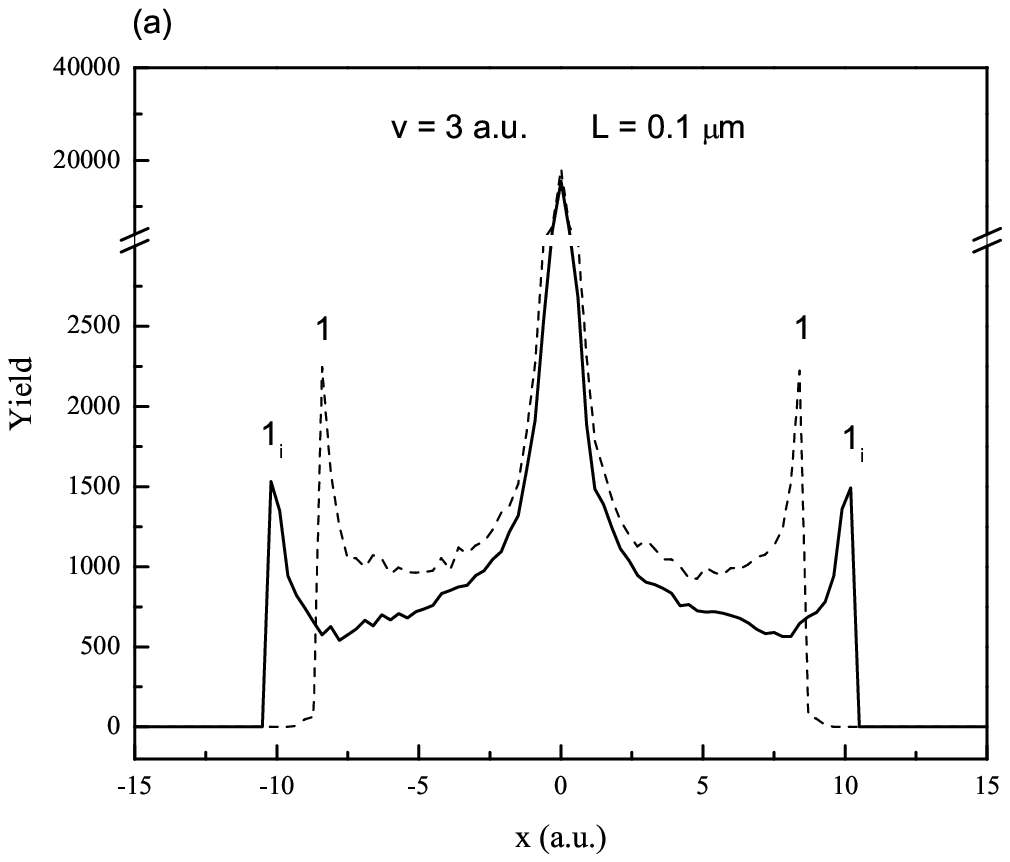}
\includegraphics[width=0.49\textwidth]{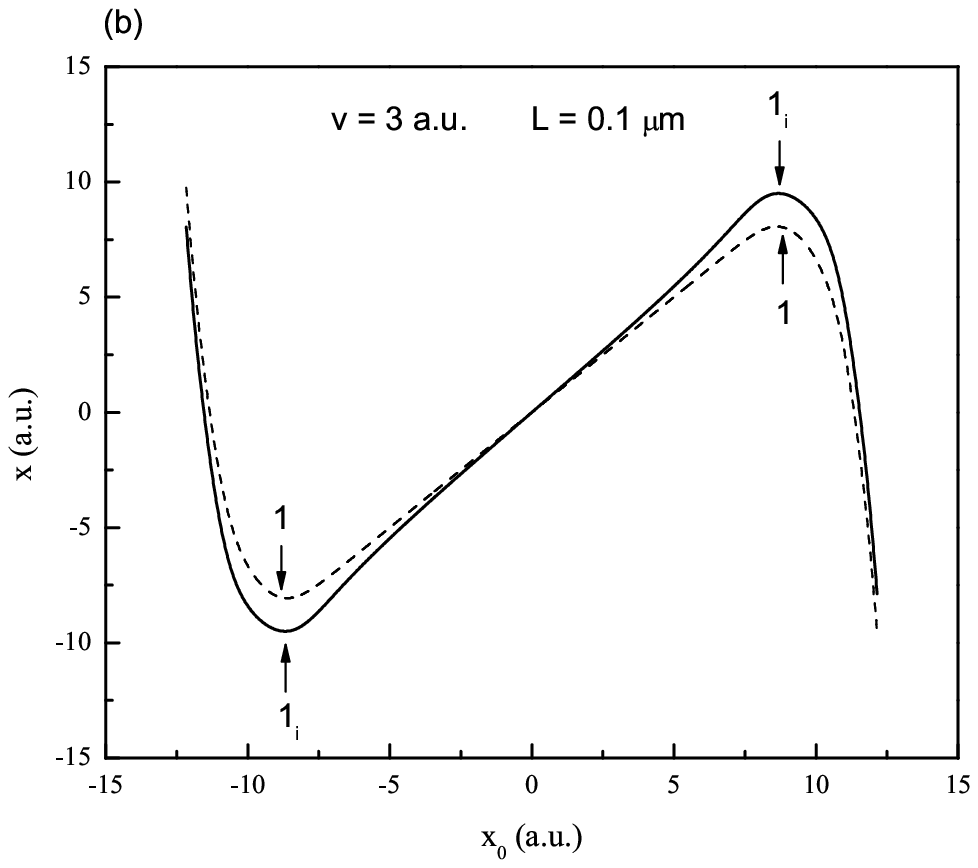}
\caption{(a) Two spatial distributions of channeled protons along
the $x$ axis in the exit transverse plane  for the proton speed of 3
a.u. and the nanotube length of 0.1 $\mu$m. The nanotube is placed
in vacuum. The dashed line corresponds to the case without the
dynamic polarization effect and the solid line to the case with it.
The size of a bin along the $x$ axis is 0.3 a.u. (b) Two mappings of
the $x_0$ axis in the entrance transverse plane to the $x$ axis in
the exit transverse plane corresponding to the two spatial
distributions.}
\label{fig01}
\end{figure}

Figure \ref{fig01}(b) shows two mappings of the $x_0$ axis in the
entrance transverse plane to the $x$ axis in the exit transverse
plane corresponding to the two spatial distributions of channeled
protons shown in Fig. \ref{fig01}(a). One can see that in each case
the mapping has two extrema, a maximum and a minimum. The extrema in
the mapping without the image interaction potential are designated
by 1 and the extrema in the mapping with it by 1$_i$. In each case
the ordinates of the extrema coincide with the abscissae of the
maxima in the corresponding spatial distribution. Since the extrema
in the mappings are the rainbow extrema, the maxima in the spatial
distributions are in fact the rainbow singularities.

If the spatial distributions of channeled protons shown in Fig.
\ref{fig01}(a) are compared with the corresponding angular
distributions of channeled protons \cite{borka06}, obtained for the
proton speed $v$ = 3 a.u. and the nanotube length $L$ = 0.1 $\mu$m,
it is seen that in the former case the rainbow effect is much more
pronounced. In this case the rainbow maxima appear even when the
image force is not included, unlike in the latter case.

\begin{figure}
\centering
\includegraphics[width=0.49\textwidth]{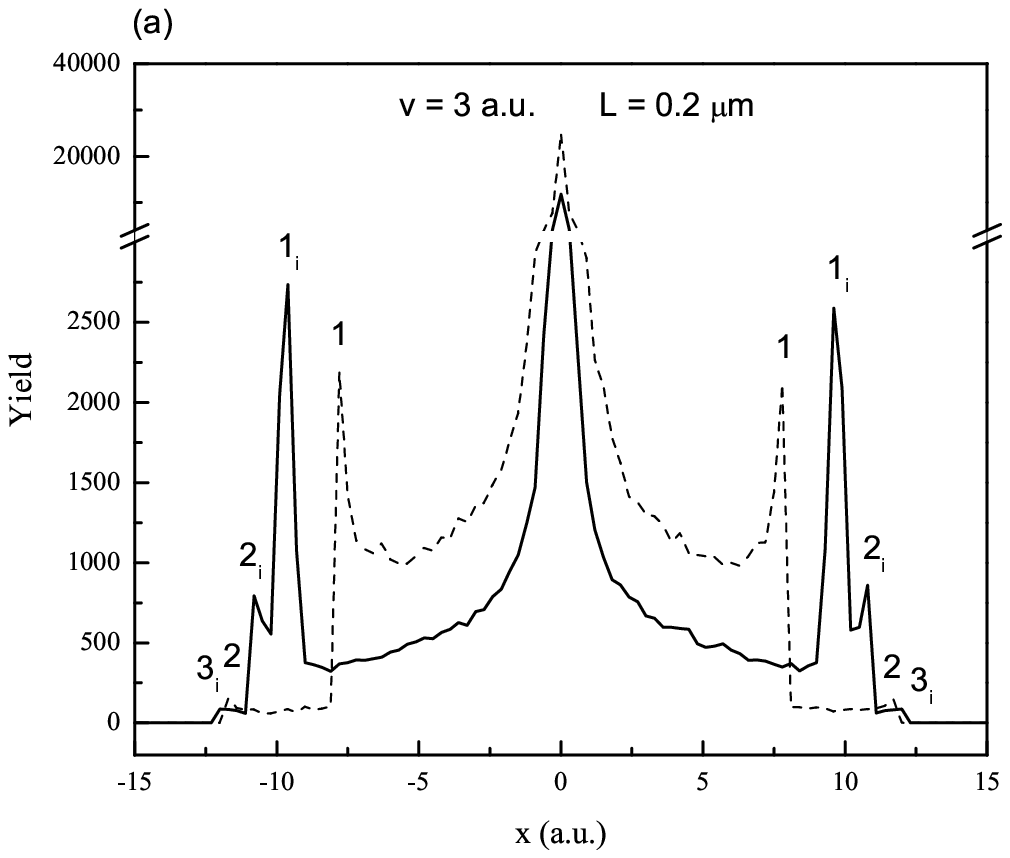}
\includegraphics[width=0.49\textwidth]{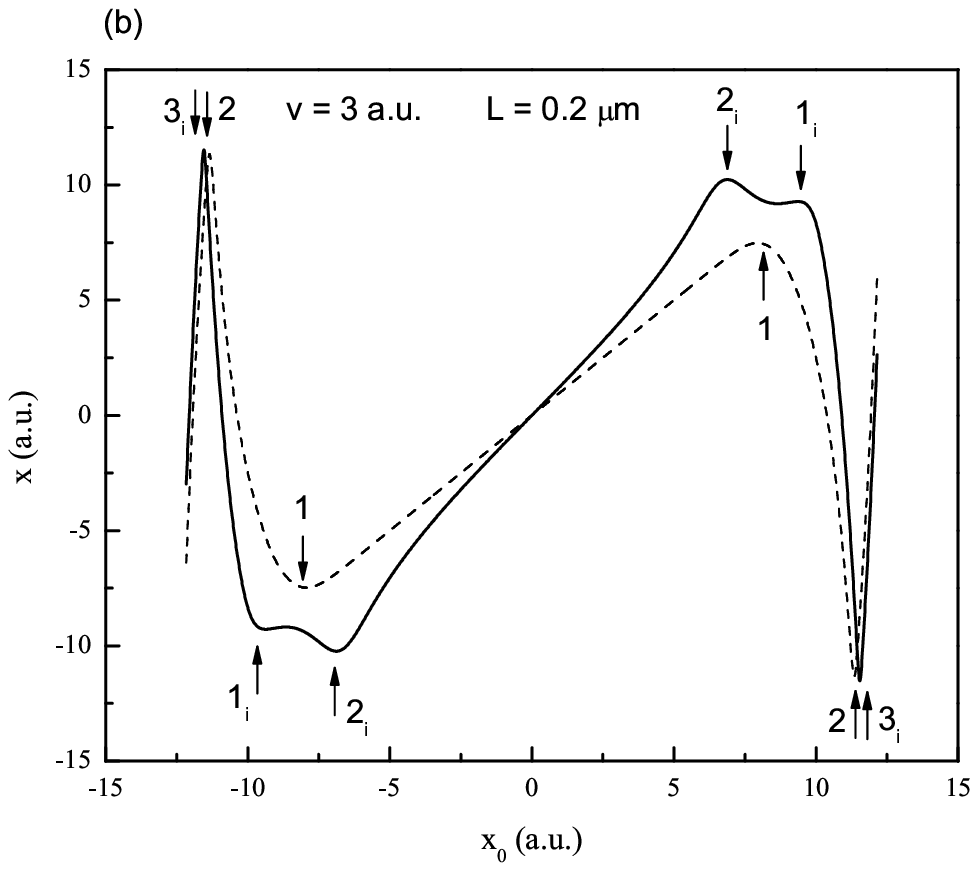}
\caption{(a) Two spatial distributions of channeled protons along
the $x$ axis in the exit transverse plane  for the proton speed of 3
a.u. and the nanotube length of 0.2 $\mu$m. The nanotube is placed
in vacuum. The dashed line corresponds to the case without the
dynamic polarization effect and the solid line to the case with it.
The size of a bin along the $x$ axis is 0.3 a.u. (b) Two mappings of
the $x_0$ axis in the entrance transverse plane to the $x$ axis in
the exit transverse plane corresponding to the two spatial
distributions.}
\label{fig02}
\end{figure}

Figure \ref{fig02}(a) shows two spatial distributions of channeled
protons along the $x$ axis in the exit transverse plane for the
proton speed $v$ = 3 a.u. and the nanotube length $L$ = 0.2 $\mu$m.
The nanotube is placed in vacuum. The spatial distributions
correspond to the cases without the image force and with it. In the
former case the spatial distribution contains a central maximum and
two pairs of peripheral maxima, designated by 1 and 2, and in the
latter case a central maximum and three pairs of peripheral maxima,
designated by 1$_i$, 2$_i$ and 3$_i$. In the case with the image
force the central maximum is narrower and about two times weaker
while the peripheral maxima are more prominent than in the case
without it. In the former case the peripheral maxima are located at
$x = \pm$7.8 and $\pm$11.7 a.u., and in the latter case at $x =
\pm$9.6, $\pm$10.8 and $\pm$12.0 a.u.

Figure \ref{fig02}(b) shows two mappings of the $x_0$ axis in the
entrance transverse plane to the $x$ axis in the exit transverse
plane corresponding to the two spatial distributions of channeled
protons shown in Fig. \ref{fig02}(a). It is clear that in the case
without the dynamic polarization effect the mapping has four
extrema, two maximum and two minima. They are designated by 1 and 2.
In the case with the dynamic polarization effect the mapping has six
extrema, three maxima and three minima. They are designated by
1$_i$, 2$_i$ and 3$_i$. In each case the ordinates of the extrema
coincide with the abscissae of the maxima in the corresponding
spatial distribution.

There are two new features in Fig. \ref{fig02} when it is compared
to Fig. \ref{fig01}. First, each of the spatial distributions of
channeled protons given in Fig. \ref{fig02}(a) contains a pair of
very weak maxima, designated by 2 in the case without the image
force and by 3$_i$ in the case with it. They correspond to the very
sharp maxima and minima in the mappings given in Fig.
\ref{fig02}(b), designated by 2 and 3$_i$, which lie near the
nanotube wall. It is noteworthy that the image force does not change
much the positions and shapes of the very weak maxima in the spatial
distribution and the very sharp maxima and minima in the mapping.
This can be attributed to the dominance of the repulsive interaction
potential in the region near the nanotube wall. Second, the spatial
distribution given in Fig. \ref{fig02}(a) without the image force
included contains one pair of prominent peripheral maxima,
designated by 1, like the corresponding spatial distribution given
in Fig. \ref{fig01}(a). On the other hand, the spatial distribution
given in Fig. 2(a) with the image force included contains two pairs
of prominent peripheral maxima, designated by 1$_i$ and 2$_i$,
unlike the spatial distribution given in Fig. \ref{fig01}(a) with
the image force included and the spatial distribution given in Fig.
\ref{fig02}(a) without the image force included. They correspond to
the shallow maxima and minima in the mapping given in Fig.
\ref{fig02}(b) with the image force included, designated by 1$_i$
and 2$_i$, which lie away from the nanotube wall. These maxima
demonstrate the presence of the image force.

\begin{figure}
\centering
\includegraphics[width=0.49\textwidth]{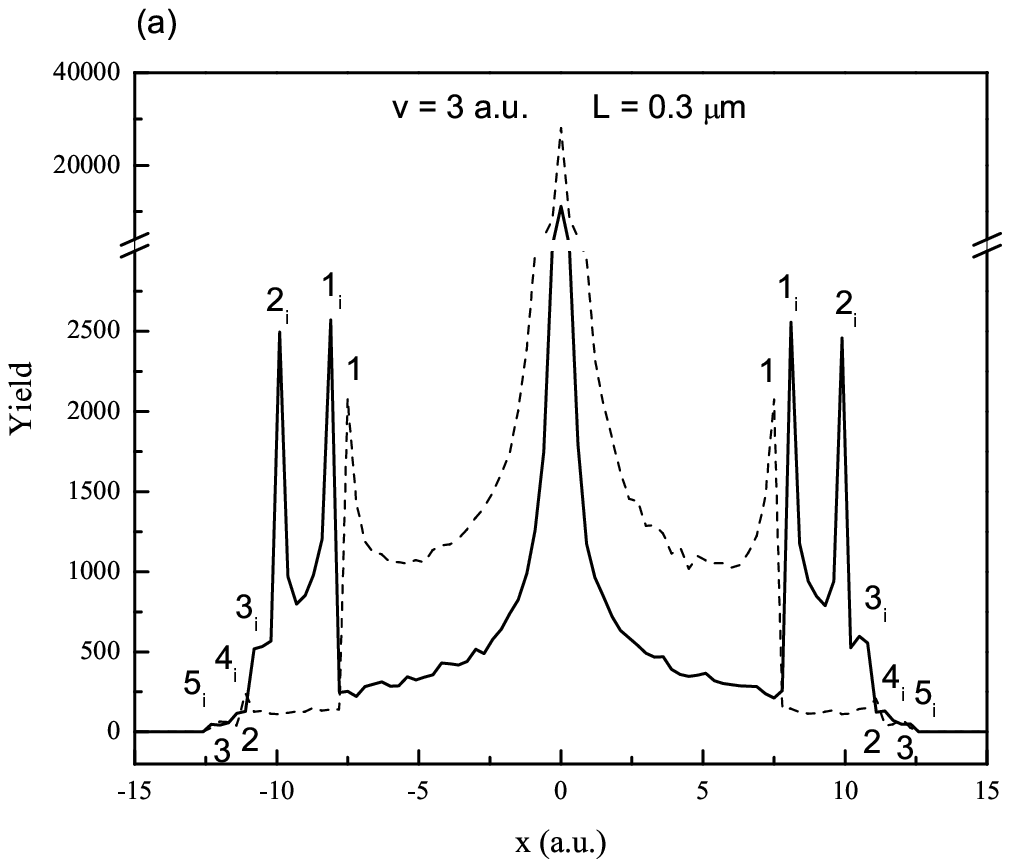}
\includegraphics[width=0.49\textwidth]{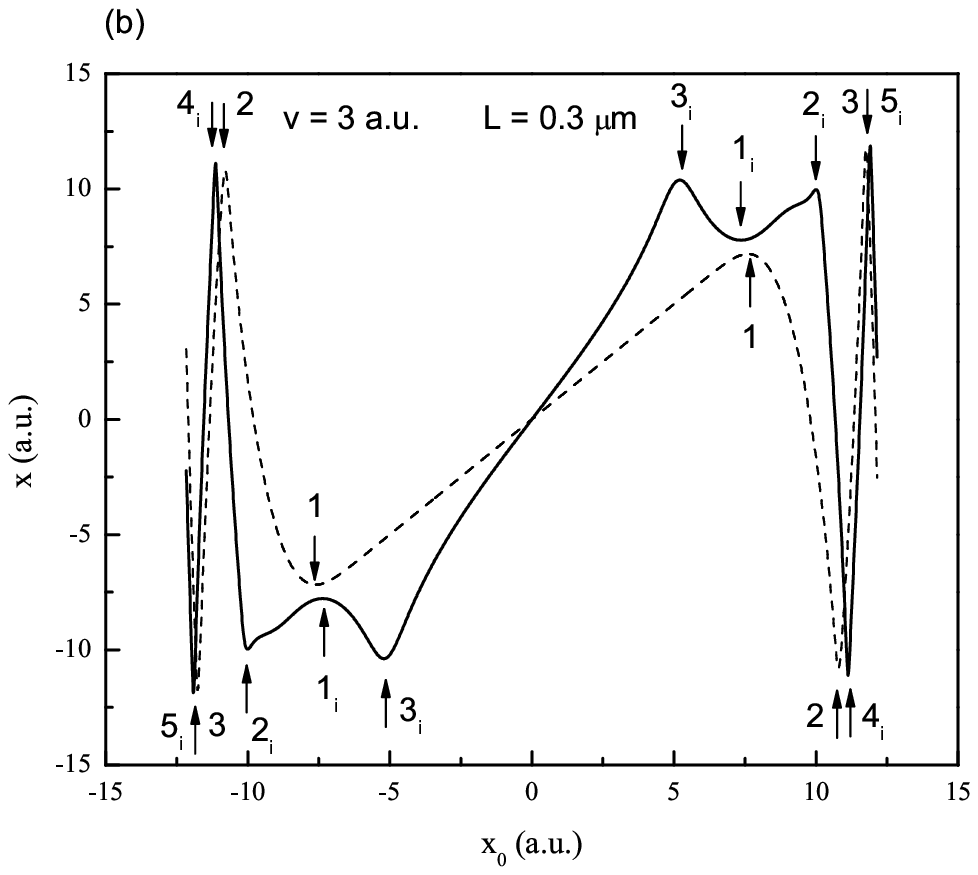}
\caption{(a) Two spatial distributions of channeled protons along
the $x$ axis in the exit transverse plane  for the proton speed of 3
a.u. and the nanotube length of 0.3 $\mu$m. The nanotube is placed
in vacuum. The dashed line corresponds to the case without the
dynamic polarization effect and the solid line to the case with it.
The size of a bin along the $x$ axis is 0.3 a.u. (b) Two mappings of
the $x_0$ axis in the entrance transverse plane to the $x$ axis in
the exit transverse plane corresponding to the two spatial
distributions.}
\label{fig03}
\end{figure}

Figure \ref{fig03}(a) shows two spatial distributions of channeled
protons along the $x$ axis in the exit transverse plane for the
proton velocity $v$ = 3 a.u. and the nanotube length $L$ = 0.3
$\mu$m. The nanotube is placed in vacuum. The spatial distributions
correspond to the cases without the image interaction potential and
with it. In the former case the spatial distribution contains a
central maximum and three pairs of peripheral maxima, designated by
1, 2 and 3, and in the latter case a central maximum and five pairs
of peripheral maxima, designated by 1$_i$, 2$_i$, 3$_i$, 4$_i$ and
5$_i$. The maxima designated by 2, 3, 4$_i$ and 5$_i$ are very weak.
In the case with the image interaction potential the central maximum
is much narrower and about three times weaker while the peripheral
maxima are much more prominent than in the case without it. In the
former case the peripheral maxima are located at $x = \pm$7.5,
$\pm$11.1 and $\pm$12.1 a.u., and in the latter case at $x =
\pm$8.1, $\pm$9.9, $\pm$10.7, $\pm$11.4 and $\pm$12.2 a.u.

Figure \ref{fig03}(b) shows two mappings of the $x_0$ axis in the
entrance transverse plane to the $x$ axis in the exit transverse
plane corresponding to the two spatial distributions of channeled
protons shown in Fig. \ref{fig03}(a). It is evident that in the case
without the image force the mapping has six extrema, three maxima
and three minima. They are designated by 1, 2 and 3. In the case
with the image force the mapping has 10 extrema, five maxima and
five minima. They are designated by 1$_i$, 2$_i$, 3$_i$, 4$_i$ and
5$_i$. The maxima and minima designated by 2, 3, 4$_i$ and 5$_i$ are
very sharp and lie near the nanotube wall.

\begin{figure}
\centering
\includegraphics[width=0.49\textwidth]{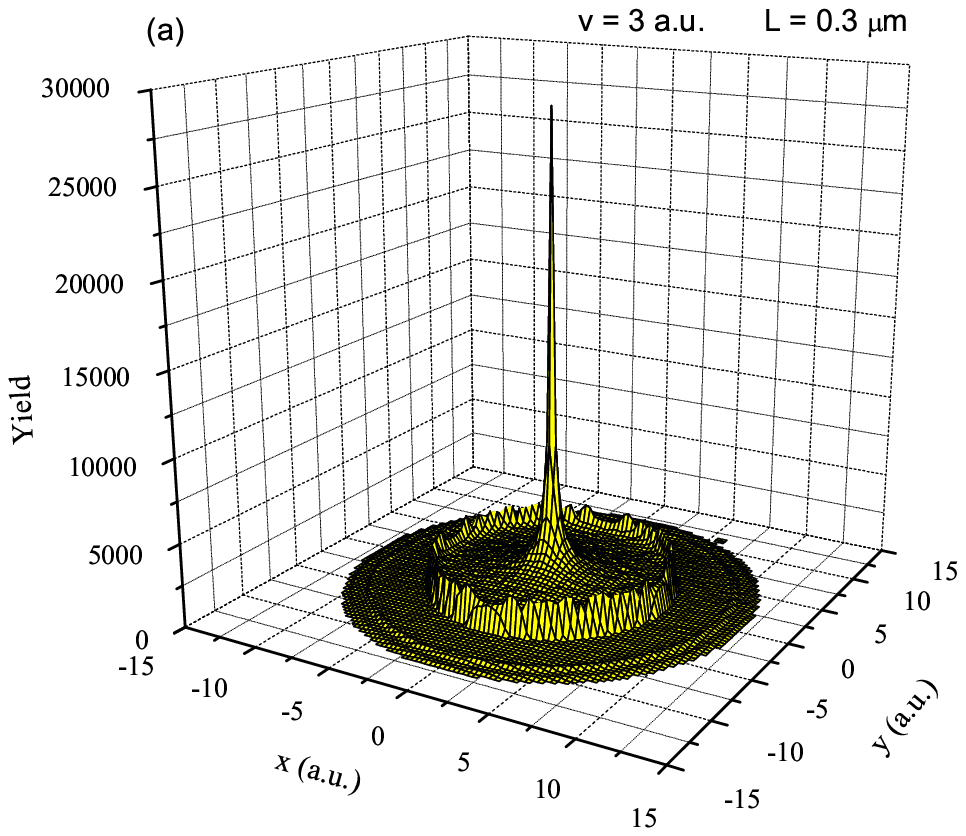}
\includegraphics[width=0.49\textwidth]{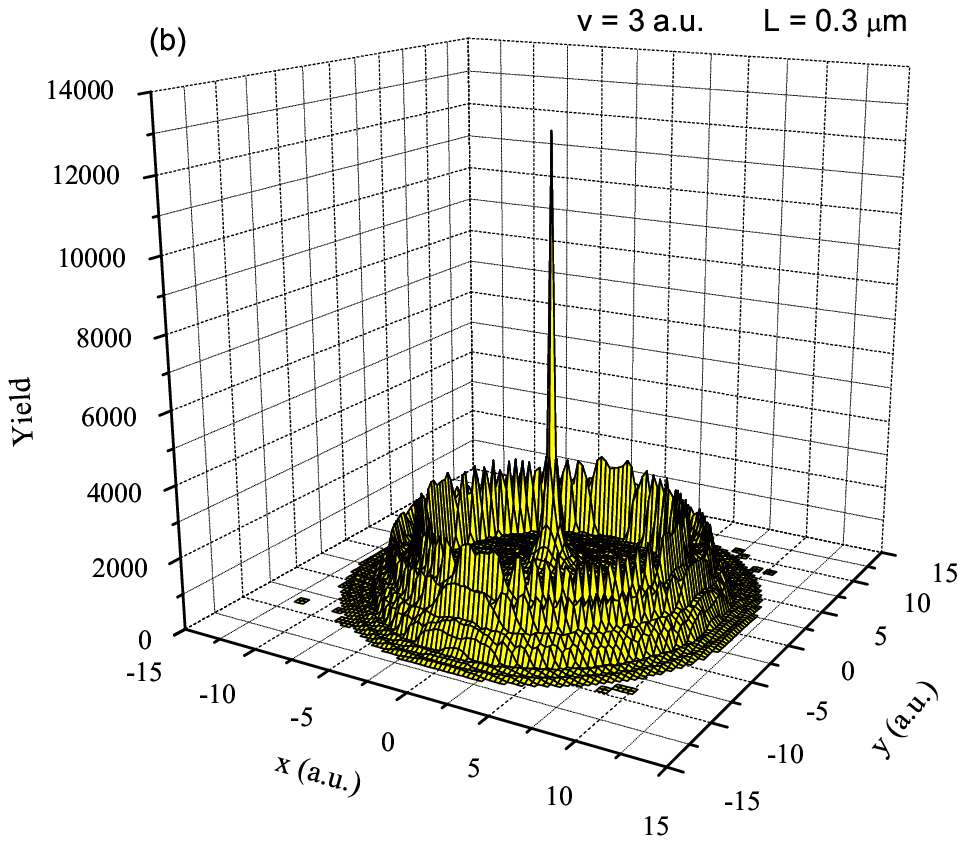}
\caption{Two spatial distributions of channeled protons in the exit
transverse plane for the proton speed of 3 a.u. and the nanotube
length of 0.3 $\mu$m. The nanotube is placed in vacuum. The spatial
distributions correspond to the cases (a) without the dynamic
polarization effect and (b) with it. The sizes of a bin along the
$x$ and $y$ axes are 0.3 a.u.} \label{fig04}
\end{figure}

Figure \ref{fig04} shows the spatial distributions of channeled
protons in the exit transverse plane for the proton speed $v$ = 3
a.u. and the nanotube length $L$ = 0.3 $\mu$m without the dynamic
polarization effect and with it. The nanotube is placed in vacuum.
The weak azimuthal asymmetry of the spatial distributions would be
less pronounced if the initial number of protons were larger.

It is obvious from Figs. \ref{fig01}-\ref{fig03} that for a longer
nanotube the image force makes the spatial distributions of
channeled protons richer. Figure \ref{fig04} shows clearly that the
image force causes a significant increase of the proton flux near
the nanotube wall, in a similar way as it was observed by Zhou et
al. \cite{zhou05}. It is important to note that this increase of the
proton flux occurs at the distances from the nanotube wall of the
order of 0.2 nm, coinciding with the typical separations between the
nanotube wall and the atoms that can be adsorbed on them. Therefore,
it is evident that the image force makes possible the use of proton
channeling for detecting and locating the atoms and molecules
adsorbed on the nanotube wall.

\begin{figure}
\centering
\includegraphics[width=0.49\textwidth]{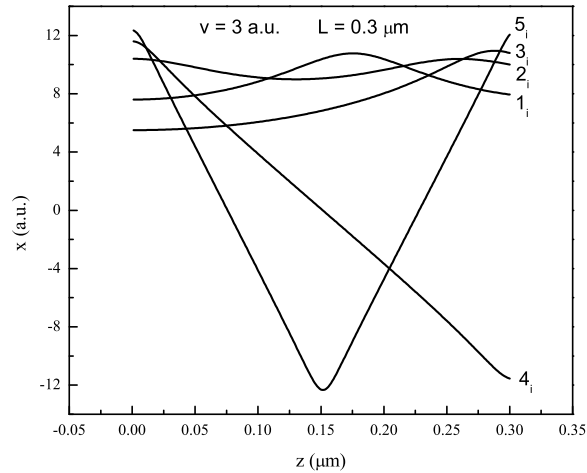}
\caption{Five proton trajectories in the $xz$ plane for the $x_0$
coordinates of the initial proton positions corresponding to the
five maxima in the spatial distribution of channeled protons shown
in Fig. 3(a) with the dynamic polarization effect included.}
\label{fig05}
\end{figure}

Figure \ref{fig05} gives five proton trajectories in the $xz$ plane
for the $x_0$ coordinates of the initial proton position
corresponding to the five maxima in the spatial distribution of
channeled protons given in Fig. \ref{fig03}(a) with the image
interaction potential. The analysis shows that these trajectories,
which are called the rainbow trajectories, can be classified
according to the number of proton deflections within the total
interaction potential well \cite{borka06}. The number of proton
deflections determines the order of the rainbow. One can see that
trajectory 4$_i$, corresponding to maximum 4$_i$ in the spatial
distribution, includes one proton deflection. Therefore, the rainbow
it represents is the primary rainbow. The rainbows represented by
trajectories 1$_i$, 3$_i$ and 5$_i$, corresponding to maxima 1$_i$,
3$_i$ and 5$_i$ in the spatial distribution, are the secondary
rainbows, and that represented by trajectory 2$_i$, corresponding to
maxima 2$_i$ in the spatial distribution, is the tertiary rainbow.

We next compare the effects of the image force on the spatial
distributions of protons channeled in the nanotube placed in vacuum
and embedded in SiO$_2$. The proton speed is increased to $v$ = 5
a.u. and the nanotube lengths are $L$ = 0.3 and 0.5 $\mu$m.

\begin{figure}
\centering
\includegraphics[width=0.49\textwidth]{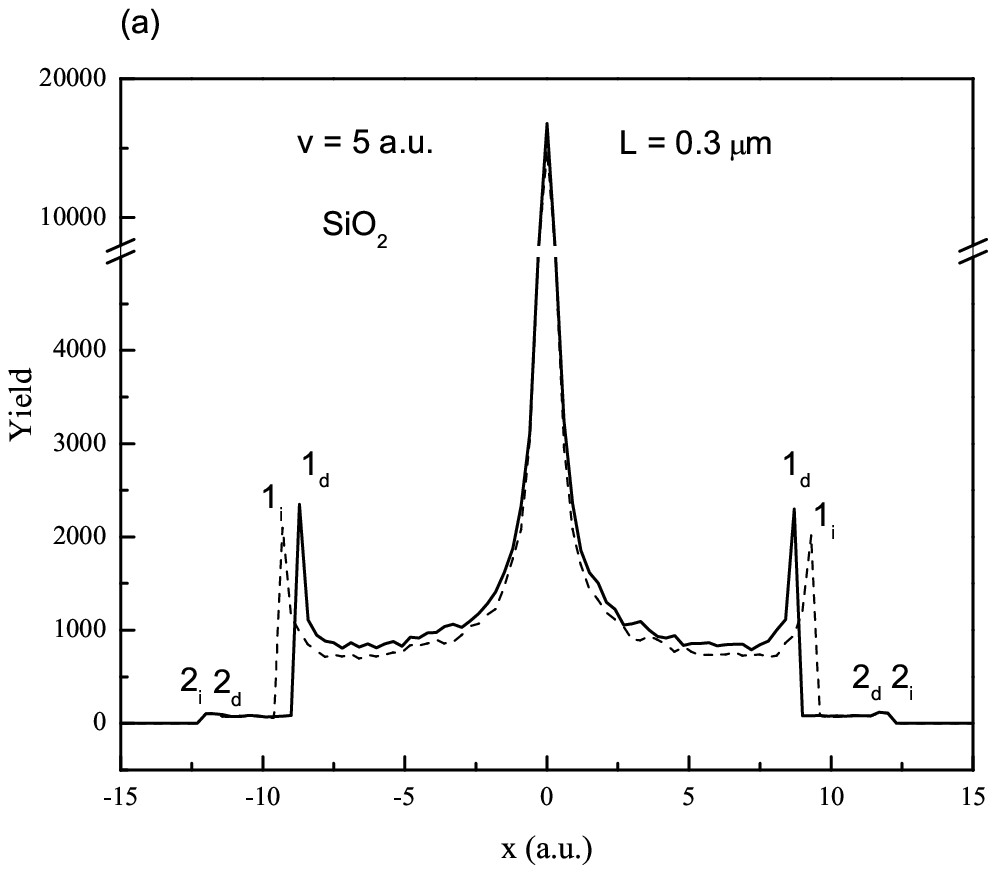}
\includegraphics[width=0.49\textwidth]{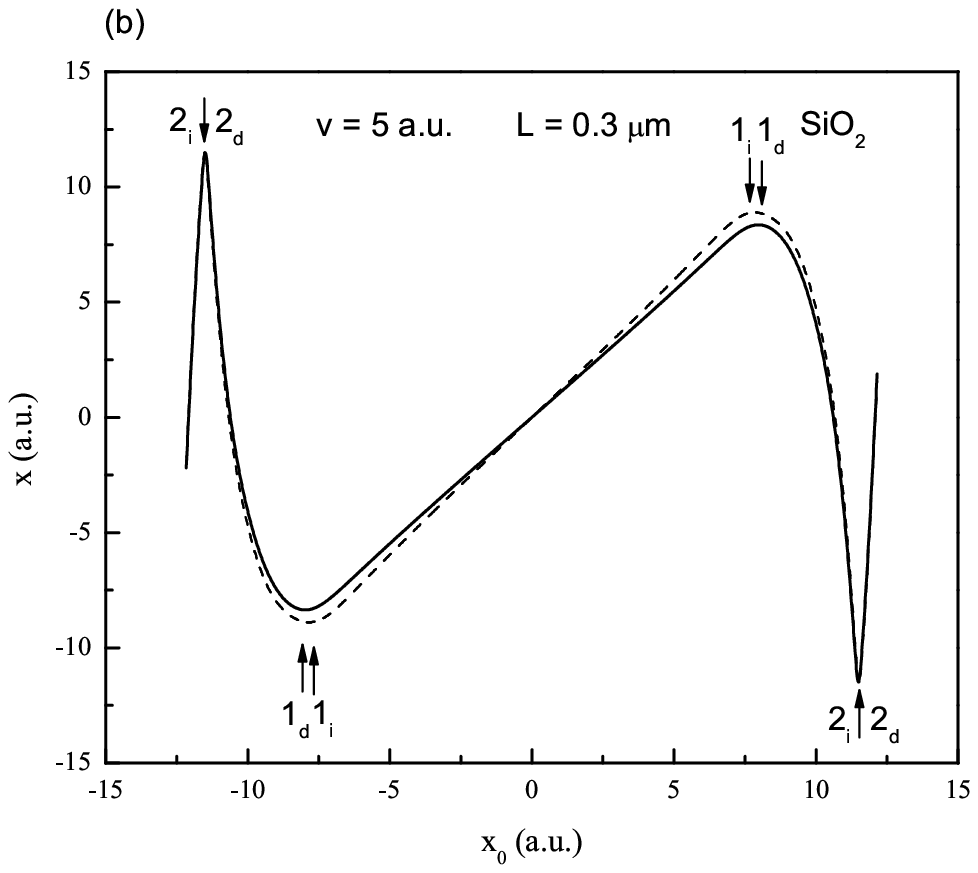}
\caption{(a) Two spatial distributions of channeled protons along
the $x$ axis in the exit transverse plane  for the proton speed of 5
a.u. and the nanotube length of 0.3 $\mu$m. The dashed line
corresponds to the case in which the nanotube is placed in vacuum
and the solid line to the case in which it is embedded in SiO$_2$.
The size of a bin along the $x$ axis is 0.3 a.u. (b) Two mappings of
the $x_0$ axis in the entrance transverse plane to the $x$ axis in
the exit transverse plane corresponding to the two spatial
distributions.}
\label{fig06}
\end{figure}

Figure \ref{fig06}(a) shows two spatial distributions of channeled
protons along the $x$ axis in the exit transverse plane for the
proton speed $v$ = 5 a.u. and the nanotube length $L$ = 0.3 $\mu$m.
They correspond to the cases in which the nanotube is placed in
vacuum and embedded in SiO$_2$. The image force is included. In each
case the spatial distribution contains a central maximum and two
pairs of peripheral maxima, designated by 1$_i$ and 2$_i$ in the
former case, and by 1$_d$ and 2$_d$ in the latter case. The maxima
designated by 2$_i$ and 2$_d$ are very weak. In the case in which
the nanotube is placed in vacuum the peripheral maxima are located
at $x = \pm$9.3 and $\pm$11.9 a.u., and in the case in which it is
embedded in SiO$_2$ at $x = \pm$8.7 and $\pm$11.9 a.u.

Figure \ref{fig06}(b) shows two mappings of the $x_0$ axis in the
entrance transverse plane to the $x$ axis in the exit transverse
plane corresponding to the two spatial distributions of channeled
protons shown in Fig. \ref{fig06}(a). One can see that in each case
the mapping has four extrema, two maxima and two minima. The extrema
in the mapping in the case in which the nanotube is placed in vacuum
are designated by 1$_i$ and 2$_i$, and in the case in which it is
embedded in SiO$_2$ by 1$_d$ and 2$_d$. The maxima and minima
designated by 2$_i$ and 2$_d$ are very sharp and lie near the
nanotube wall. In each case the ordinates of the extrema coincide
with the abscissae of the maxima in the corresponding spatial
distribution. Thus, the maxima in the spatial distributions are in
fact the rainbow singularities.

\begin{figure}
\centering
\includegraphics[width=0.49\textwidth]{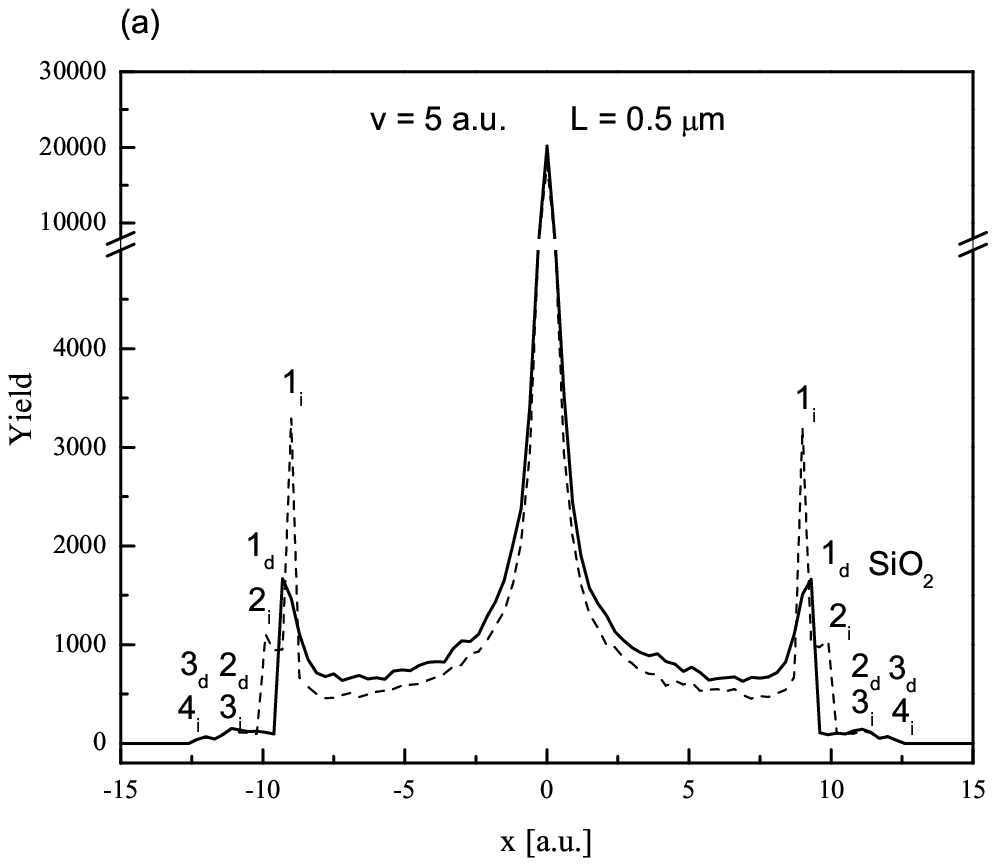}
\includegraphics[width=0.49\textwidth]{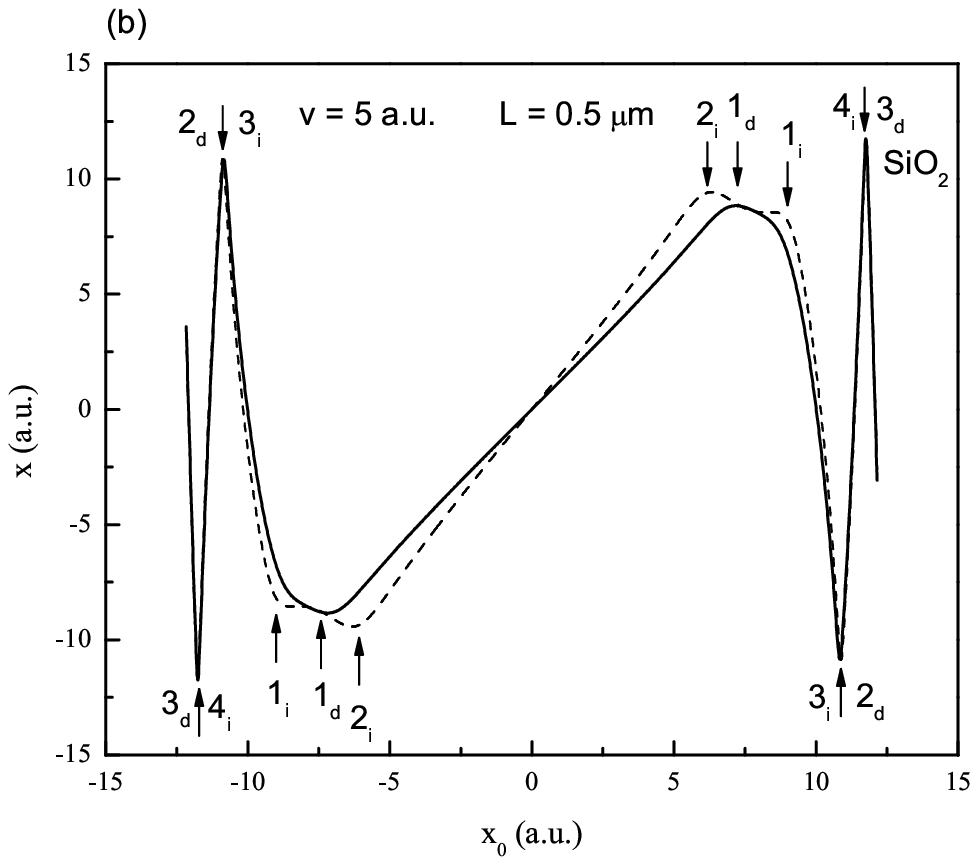}
\caption{(a) Two spatial distributions of channeled protons along
the $x$ axis in the exit transverse plane  for the proton speed of 5
a.u. and the nanotube length of 0.5 $\mu$m. The dashed line
corresponds to the case in which the nanotube is placed in vacuum
and the solid line to the case in which it is embedded in SiO$_2$.
The   size of a bin along the $x$ axis is 0.3 a.u. (b) Two mappings
of the $x_0$ axis in the entrance transverse plane to the $x$ axis
in the exit transverse plane corresponding to the two spatial
distributions of channeled protons.}
\label{fig07}
\end{figure}

Figure \ref{fig07}(a) shows two spatial distributions of channeled
protons along the $x$ axis in the exit transverse plane for the
proton speed $v = $5 a.u. and the nanotube length $L$ = 0.5 $\mu$m.
They correspond to the cases in which the nanotube is placed in
vacuum and embedded in SiO$_2$. The dynamic polarization effect is
included. In the former case the spatial distribution contains a
central maximum and four pairs of peripheral maxima, designated by
1$_i$, 2$_i$, 3$_i$ and 4$_i$, and in the latter case a central
maximum and three pairs of peripheral maxima, designated by 1$_d$,
2$_d$ and 3$_d$. The maxima designated by 3$_i$, 4$_i$, 2$_d$ and
3$_d$ are very weak. In the case in which the nanotube is placed in
vacuum the peripheral maxima are located at $x = \pm$9.0, $\pm$9.9,
$\pm$11.2 and $\pm$12.1 a.u., and in the case in which it is
embedded in SiO$_2$ at $x = \pm$9.3, $\pm$11.2 and $\pm$12.1 a.u.

Figure \ref{fig07}(b) shows two mappings of the $x_0$ axis in the
entrance transverse plane to the $x$ axis in the exit transverse
plane corresponding to the two spatial distributions of channeled
protons shown in Fig. \ref{fig07}(a). It is evident that in the case
in which the nanotube is placed in vacuum the mapping has eight
extrema, four maxima and four minima. They are designated by 1$_i$,
2$_i$, 3$_i$ and 4$_i$. In the case in which the nanotube is
embedded in SiO$_2$ the mapping has six extrema, three maxima and
three minima. They are designated by 1$_d$, 2$_d$ and 3$_d$. The
maxima and minima designated by 3$_i$, 4$_i$, 2$_d$ and 3$_d$ are
very sharp and lie near the nanotube wall. In each case the
ordinates of the extrema coincide with the abscissae of the maxima
in the corresponding spatial distribution.

It is interesting to note that the spatial distribution of channeled
protons given in Fig. \ref{fig07}(a) when nanotube is embedded in
SiO$_2$ contains one pair of prominent peripheral maxima rather than
two such pairs of maxima, as in the other spatial distribution given
in this figure (when nanotube is placed in vacuum). This is
explained by the weakening of the image force at this proton speed
due to the presence of the dielectric medium, as it was shown by
Borka et al. \cite{borka08}.

\begin{figure}
\centering
\includegraphics[width=0.49\textwidth]{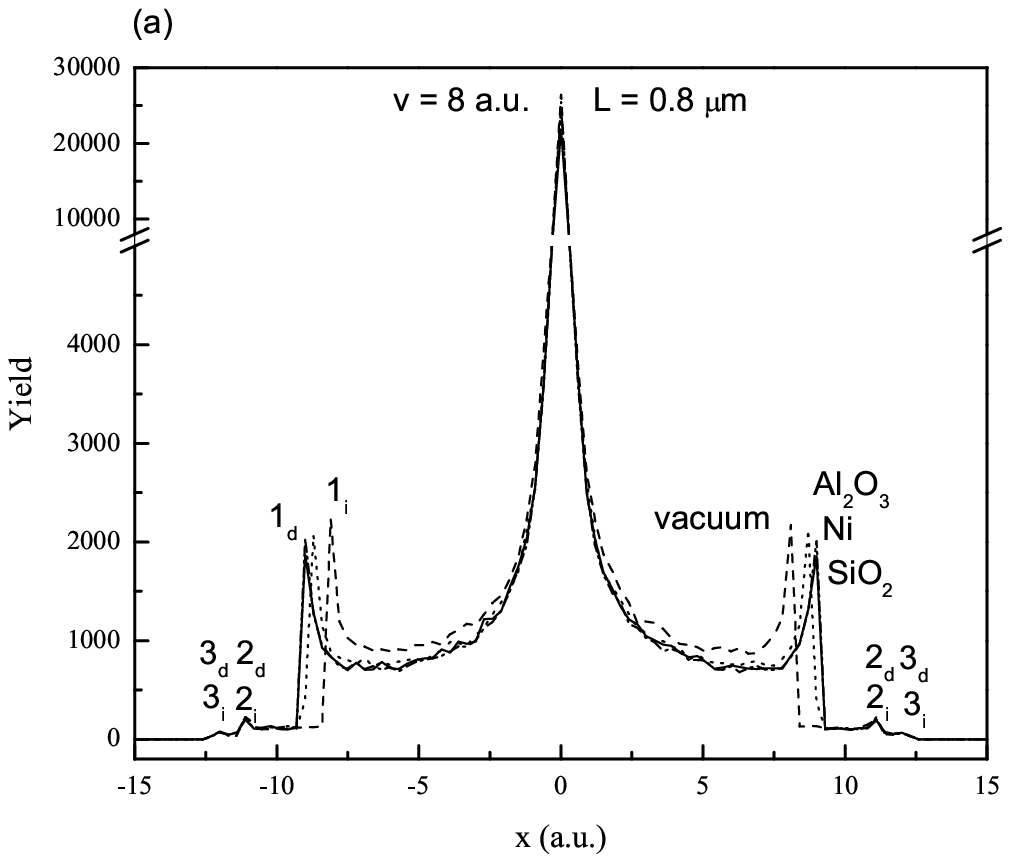}
\includegraphics[width=0.49\textwidth]{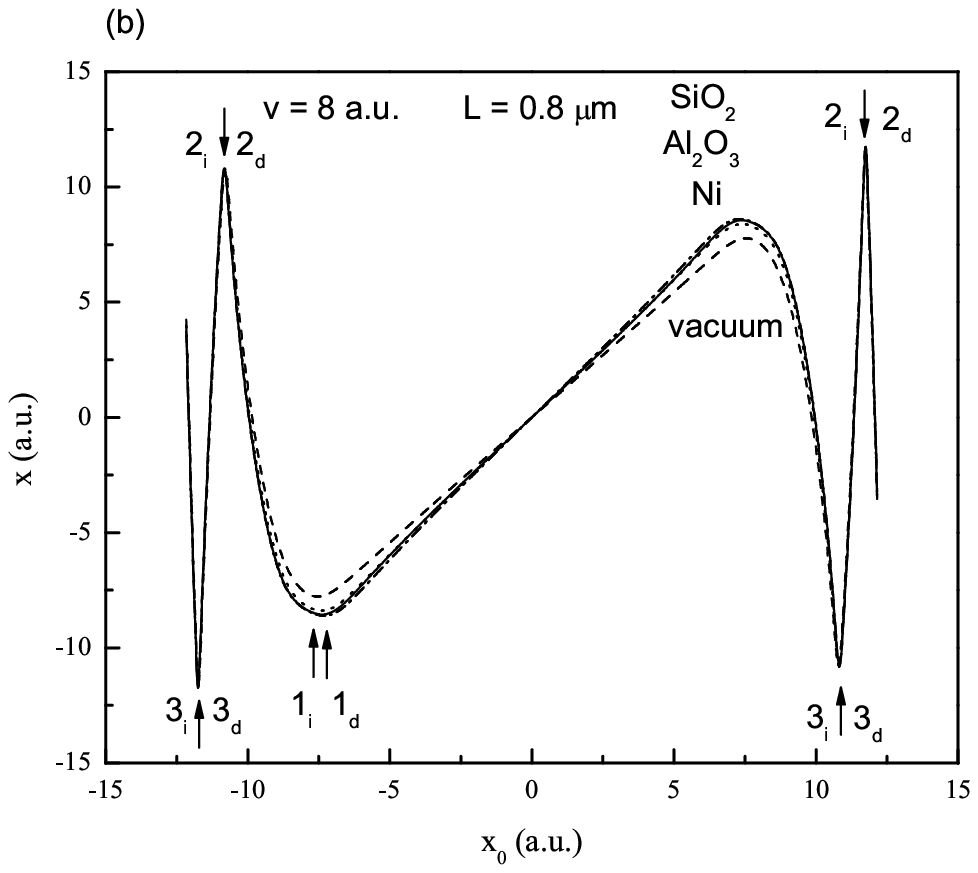}
\caption{(a) Four spatial distributions of channeled protons along
the $x$ axis in the exit transverse plane for the proton speed of 8
a.u. and the nanotube length of 0.8 $\mu$m. The dashed line
corresponds to the case in which the nanotube is placed in vacuum,
and the solid, dotted and dash-dotted lines correspond to the cases
in which the nanotube is embedded in SiO$_2$, Al$_2$O$_3$ and Ni,
respectively. The size of a bin along the $x$ axis is 0.3 a.u. (b)
Four mappings of the $x_0$ axis in the entrance transverse plane to
the $x$ axis in the exit transverse plane corresponding to the four
spatial distributions.}
\label{fig08}
\end{figure}

Let us now compare the spatial distributions of channeled protons
along the $x$ axis in the exit transverse plane for the proton speed
$v$ = 3 a.u. and the nanotube length $L$ = 0.3 $\mu$m, for $v$ = 5
a.u. and $L$ = 0.5 $\mu$m, and for $v$ = 8 a.u. and $L$ = 0.8
$\mu$m. The dynamic polarization effect is included. These spatial
distributions are characterized by the same duration of the process
of proton channeling, i.e., by the same proton dwell time. In the
case in which $v$ = 3 a.u. and $L$ = 0.3 $\mu$m the spatial
distribution with the nanotube placed in vacuum contains five
rainbow maxima [see Fig. \ref{fig03}(a)], as the spatial
distribution with the nanotube placed in SiO$_2$. In the case in
which $v$ = 5 a.u. and $L$ = 0.5 $\mu$m the spatial distribution
with the nanotube placed in vacuum contains four rainbow maxima, and
the spatial distribution with the nanotube placed in SiO$_2$ three
rainbow maxima [see Fig. \ref{fig07}(a)]. In the case in which $v$ =
8 a.u. and $L$ = 0.8 $\mu$m the spatial distribution with the
nanotube placed in vacuum contains three rainbow maxima, as the
spatial distribution with the nanotube placed in SiO$_2$ [see Fig.
\ref{fig08}(a)]. It is clear that for the same proton dwell time the
number of rainbow maxima in the spatial distribution decreases with
the proton speed.

Figure \ref{fig08}(a) shows four spatial distributions of channeled
protons along the $x$ axis in the exit transverse plane for the
proton speed $v$ = 8 a.u. and the nanotube length $L$ = 0.8 $\mu$m.
They correspond to the cases in which the nanotube is placed in
vacuum and embedded in SiO$_2$, Al$_2$O$_3$ and Ni. The image
interaction potential is included. In each case the spatial
distribution contains a strong central maximum and three pairs of
peripheral maxima, designated by 1$_i$, 2$_i$ and 3$_i$ in the case
in which the nanotube is placed in vacuum, and by 1$_d$, 2$_d$ and
3$_d$ in the cases in which it is embedded in SiO$_2$, Al$_2$O$_3$
and Ni. It is clear that the spatial distributions in the cases in
which the nanotube is embedded in the three dielectric media are
very similar to each other, and that they do not differ
significantly from the spatial distribution in the case in which it
is placed in vacuum. The spatial distributions in the cases in which
the nanotube is embedded in SiO$_2$ and Ni almost coincide.

Figure \ref{fig08}(b) shows four mappings of the $x_0$ axis in the
entrance transverse plane to the $x$ axis in the exit transverse
plane corresponding to the four spatial distributions of channeled
protons shown in Fig. \ref{fig08}(a). One can see that in each case
the mapping has six extrema, three maxima and three minima. The
extrema in the mapping in the case in which the nanotube is placed
in vacuum are designated by 1, 2$_i$ and 3$_i$, and in the cases in
which it is embedded in SiO$_2$, Al$_2$O$_3$ and Ni by 1$_d$, 2$_d$
and 3$_d$.

Let us now consider two different types of proton trajectories in
the channeling process under consideration. One should be aware of
the facts that in the case without the attractive interaction
potential the protons can oscillate transversely only between the
sides of the nanotube wall, while in the case with it they can
oscillate transversely between the sides of the nanotube wall and
also around the two minima of the total interaction potential, which
appear due to the presence of the attractive interaction potential.

\begin{figure}
\centering
\includegraphics[width=0.49\textwidth]{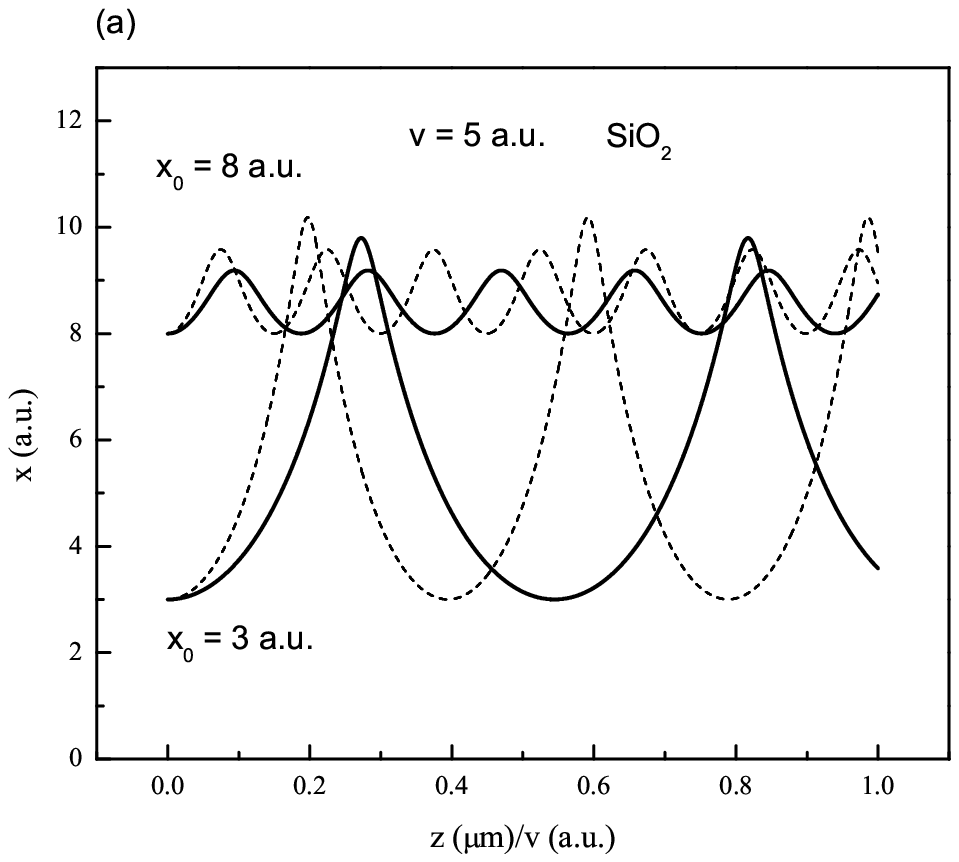}
\includegraphics[width=0.49\textwidth]{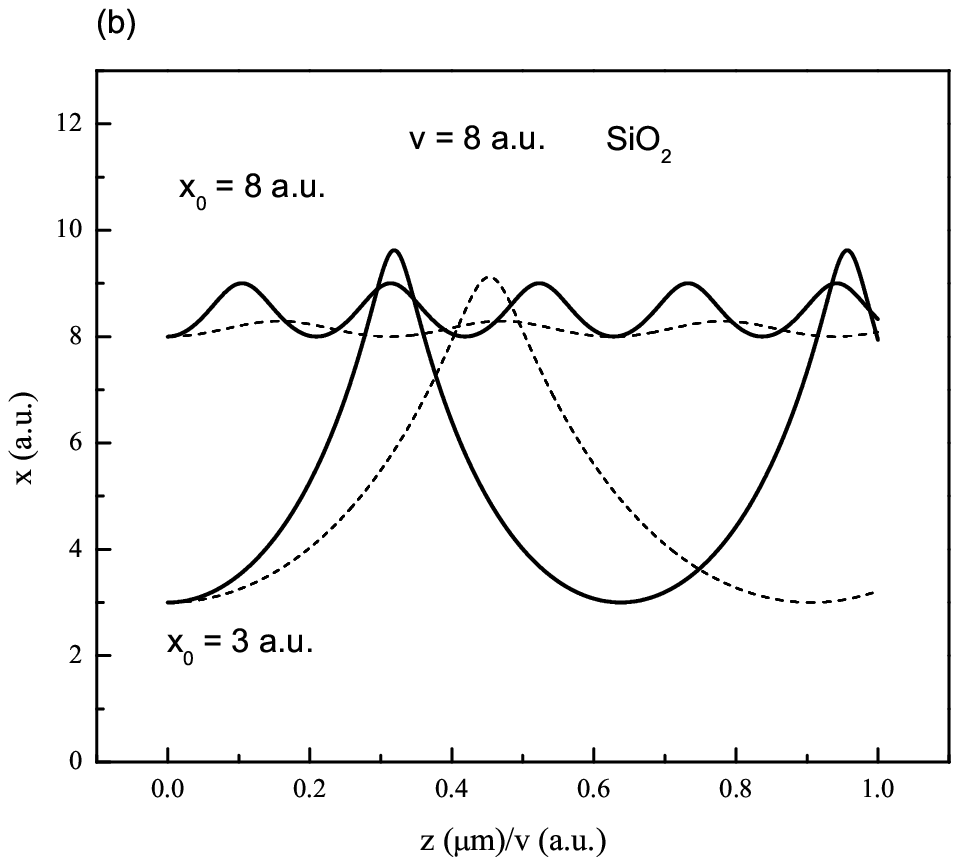}
\caption{Eight proton trajectories in the $x(z/v)$ plane for the
$x_0$ coordinates of the initial proton positions of 3 and 8 a.u.
and the proton speeds of (a) 5 a.u. and (b) 8 a.u. The dashed lines
correspond to the cases in which the nanotube is placed in vacuum
and the solid lines to the cases in which it is embedded in
SiO$_2$.}
\label{fig09}
\end{figure}

Figure \ref{fig09} shows the proton trajectories in the $x(z/v)$
plane for the $x_0$ coordinates of the initial proton positions of 3
and 8 a.u. and the proton speeds $v$ = 5 and 8 a.u. They correspond
to the cases in which the nanotube is placed in vacuum and embedded
in SiO$_2$. The image force is included. The values of $|x_0|$ are
sufficiently small, i.e., smaller than about 11 a.u., to prevent the
transverse proton oscillations between the sides of the nanotube
wall. Instead, the protons oscillate transversely around the two
minima appearing due to the presence of the attractive interaction
potential, having the $x$ coordinates about $\pm$9.3 a.u. It is
obvious that the amplitudes of the transverse proton oscillations
are smaller and their periods larger for the larger proton speed
($v$ = 8 a.u.). This is attributed to the weakening of the
attractive interaction potential with the proton speed. It is also
clear that for these values of $|x_0|$ the influence of the
dielectric medium on the proton trajectories is significant. For the
smaller proton speed ($v$ = 5 a.u.) the presence of the dielectric
medium causes an increase of the period of transverse proton
oscillations while for the larger proton speed ($v$ = 8 a.u.) the
situation is opposite.

\begin{figure}
\centering
\includegraphics[width=0.49\textwidth]{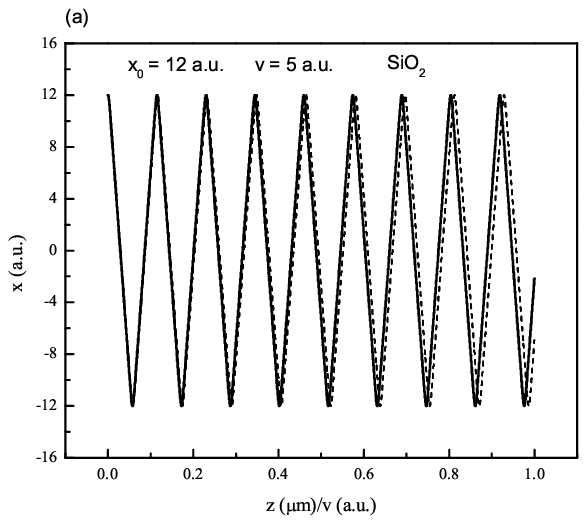}
\includegraphics[width=0.49\textwidth]{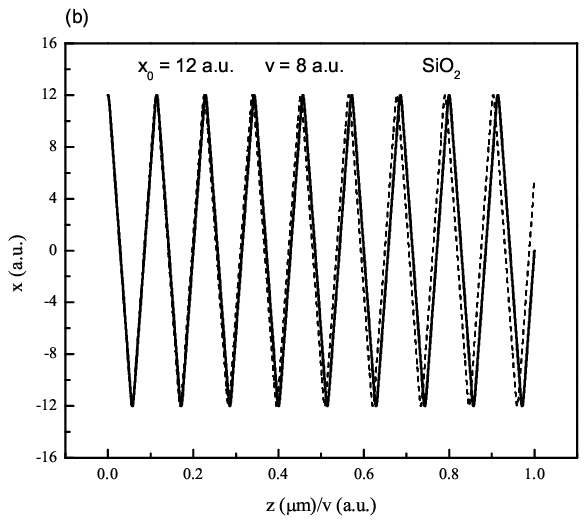}
\caption{Four proton trajectories in the $x(z/v)$ plane for the
$x_0$ coordinate of the initial proton position of 12 a.u. and the
proton speeds of (a) 5 a.u. and (b) 8 a.u. The dashed lines
correspond to the cases in which the nanotube is placed in vacuum
and the solid lines to the cases in which it is embedded in
SiO$_2$.}
\label{fig10}
\end{figure}

Figures \ref{fig10}(a) and (b) show the proton trajectories in the
$x(z/v)$ plane for the $x_0$ coordinate of the initial proton
position of 12 a.u. and the proton speeds $v$ = 5 and 8 a.u. They
correspond to the cases in which the nanotube is placed in vacuum
and embedded in SiO$_2$. The dynamic polarization effect is
included. The value of  $|x_0|$ is sufficiently large, i.e., larger
than about 11 a.u., to enable the transverse proton oscillations
between the sides of the nanotube wall. Since the nanotube wall is
defined by the repulsive interaction potential of the Doyle-Turner
type, the proton trajectories resemble those in the case of a box
with rigid walls. It is also evident that for this value of $|x_0|$
the influence of the dielectric medium on the proton trajectories is
very small, even after a number of transverse proton oscillations
between the nanotube walls.

\section{Conclusions}

We have presented theoretical investigations of the effects of
dynamic polarization of the nanotube atoms valence electrons on the
spatial distributions of protons channeled through the (11, 9)
single-wall carbon nanotubes placed in vacuum and embedded in
various dielectric media, for the proton speeds between 3 and 8 a.u.
and the nanotube lengths between 0.1 and 0.8 $\mu$m. This study
complements our previous investigations of the effects of dynamic
polarization on the angular distributions of protons channeled
through the nanotubes, where we found that the image force gives
rise to the rainbow effect \cite{borka06,borka07}, which can be
strongly affected by the surrounding dielectric medium for the
proton speeds above 3 a.u. \cite{borka08}. Similarly, our present
study has revealed that the prominent rainbow maxima exist in the
spatial distributions of channeled protons, which correspond to the
well-defined extrema in the mapping of the entrance transverse plane
to the exit transverse plane. While the rainbows in the angular
distributions of channeled ions can be easily measured and applied
to provide information on the structure and atomic forces inside the
nanotubes, the rainbows in the spatial distributions of channeled
ions can be employed for detecting and locating the atoms and
molecules intercalated in the nanotubes.

Specifically, besides the strong central maxima in all the spatial
distributions of channeled protons, we have found that, as the
nanotube length increases, the very weak maxima lying near the
nanotube wall are present too, without and with the image force and
the dielectric media included. However, our most important finding
is the prominent peripheral maxima in the spatial distributions at
the distances from the nanotube wall of the order of a few tenths of
a nanometer. These distances are perfectly suitable for applying
proton channeling to probe the atoms and molecules adsorbed on the
nanotube wall. We have revealed that the image force is responsible
for the appearance of the additional prominent peripheral maxima as
well as for making them more prominent at the expense of the central
maximum, as the nanotube length increases.

Since the spatial distribution of channeled protons gives us a
detailed information about the proton flux within the nanotube, it
appears that a careful studying of the speed dependence of the image
force as well as of the properties of the nanotube and its
surrounding can help us better understand and perhaps even find a
way to induce a spatial redistribution of channeled protons towards
the nanotube wall to be used for probing the atoms and molecules
intercalated in the nanotube.

Besides such diagnostic applications, we would like to mention the
possibility of producing nanosized ion beams with the nanotubes
embedded in various dielectric media for applications in biomedical
research \cite{biry05a,biry05c,bell03}. Our findings can shed more
light on the issue of shape of the spatial distribution of channeled
ions suitable for such applications, especially if there are
concerns about the increase of the proton flux in the peripheral
region of a nanosized ion beam at the expense of the flux in its
central region.

\ack

D. B., S. P., and N. N. acknowledge the support to this work
provided by the Ministry of Science of Serbia, and D. B., D. J. M.
and Z. L. M. acknowledge the supports by NSERC and PREA. D. B. would
also like to thank Professor Giuseppe Tenti and Professor Frank
Goodman for many useful discussions.

\end{document}